\documentclass{rmf-d}
\usepackage{nopageno,rmfbib,multicol,times,epsf,amsmath,amssymb,cite}
\usepackage[T1]{fontenc} 
\usepackage[]{caption2}
\usepackage{graphicx}
\usepackage{blindtext}
\usepackage{color}
\usepackage[colorlinks=true]{hyperref}

\mathchardef\mhypen"2D
\newcommand{\vphi}{\varphi}
\newcommand{\veps}{\varepsilon}

\renewcommand{\bar}[1]{\overline{#1}}

\newcommand{\cmpq}{\widetilde{C}_{\vphi q}^{(-)33}\rule{0.0pt}{12pt}}
\newcommand{\cubr}{\widetilde{C}_{uB\,r}^{33}\rule{0.0pt}{12pt}}
\newcommand{\cubi}{\widetilde{C}_{uB\,i}^{33}\rule{0.0pt}{12pt}}
\newcommand{\cub}{\widetilde{C}_{uB}^{33}\rule{0.0pt}{12pt}}

\newcommand{\ccubr}{C_{uB\,r}^{33}\rule{0.0pt}{12pt}}
\newcommand{\ccubi}{C_{uB\,i}^{33}\rule{0.0pt}{12pt}}
\newcommand{\ccub} {C_{uB}^{33}\rule{0.0pt}{12pt}}

\newcommand{\C}{\widetilde{C}}
\renewcommand{\O}{\mathcal{O}}

\def\rmfcornisa{Research OR Education of Physics \hfill\rmf\ {\bf ??} (*?*) ???--???
\hfill MES? A\~NO?}

\def\rmfcintilla{{\it Rev.\ Mex.\ Fis.\/} {\bf ??} (*?*) (????) ???--???}
\clearpage \rmfcaptionstyle 
\begin{document}

\title{Top quark effective couplings from top-pair tagged
  photoproduction in $\boldsymbol{pe^-}$ collisions
\vspace{-6pt}}
\author{Antonio O.\ Bouzas  and F.\ Larios}

\address{Departamento de F\'{\i}sica Aplicada,
CINVESTAV-IPN \\\small Carretera Antigua a Progreso Km.\ 6, Apdo.\
Postal 73 ``Cordemex''\\\small M\'erida 97310, Yucat\'an, M\'exico}

\maketitle
%
%
\recibido{15 March 2023}{16 April 2023
\vspace{-12pt}}

\begin{abstract}
  \vspace{1em} We summarize the quantitative results of our analysis
  \cite{bou22} of top-pair photoproduction in semileptonic mode in
  $pe$ collisions at the LHeC and FCC-he. We define three
  photoproduction regions, based on the rapidity acceptance range of the
  electron tagger, that provide different degrees of sensitivity to
  top-quark effective couplings.  We focus on the $t\bar{t}\gamma$
  dipole couplings and the left-handed vector $tbW$ coupling, for
  which we determine limits at both energies in the different
  photoproduction regions. We find that the LHeC and FCC-he will yield
  tight direct bounds on top dipole moments, greatly improving on
  current direct limits from hadron colliders, and direct limits on
  the $tbW$ coupling as restrictive as those expected from the HL-LHC.
  We also consider indirect limits from $b\to s \gamma$ branching
  ratio and $CP$ asymmetry, that are well known to be very sensitive
  probes of top electromagnetic dipole moments.  \vspace{1em}
\end{abstract}
\keys{ \bf{\textit{top-quark photoproduction electron-proton-collider
}} \vspace{-8pt}}
\begin{multicols}{2}

\section{Introduction}
\label{sec:intro}

Future $pe^-$ colliders, such as the Large Hadron-electron Collider
(LHeC) and the Future Circular Collider (FCC-he), will have among
their most important areas of research the study of the top quark
effective couplings to the Higgs and the electroweak bosons
\cite{LHeC2020}. Indeed, the top quark effective couplings constitute
a phenomenological research area of great interest
\cite{yuan1,kozachuk}.  Top-pair and single-top production at
the LHeC are very good probes for charged-current (CC) $tbW$ and
neutral-current (NC) $ttZ$ effective couplings \cite{sar14,mellado}.
Also, anomalous magnetic and electric dipole moments of the top quark
can be very well probed through top-pair photoproduction in
electron-proton collisions \cite{bou13b,billur17,billur20}.

In our recent paper \cite{bou22} we obtain limits on the top-quark
anomalous electromagnetic dipole moments and its left-handed vector
$tbW$ coupling, in the context of the Standard Model Effective Field
Theory (SMEFT), by means of Monte Carlo simulations including parton
showering and hadronization, and fast detector simulation, for both the
LHeC and the FCC-he. We compute the photoproduction cross section in
tree-level QED, taking the complete kinematics into account, including
the scattered-electron transverse momentum. This allows us to
determine the phase-space region where the photoproduction process is
sensitive to the top anomalous dipole moments, and that in which it
is sensitive to the anomalous $tbW$ coupling.

Also studied in \cite{bou22} are the indirect limits on top dipole
moments from the decays $B\to X_s\gamma$. We update our previous
results \cite{bou13a} for those limits, and discuss in detail
the fact that there are currently two different sets of such limits,
based on two incompatible theoretical computations
\cite{hewett,crivellin15} of the new physics contributions to the
branching ratio for $B\to X_s\gamma$ and its associated $CP$
asymmetry.

In this note we discuss the quantitative results from \cite{bou22} on
direct limits on the anomalous top dipole moments and left-handed
vector $tbW$ coupling, and comment also on the indirect limits
obtained there. We discuss experimental limits by the CMS
collaboration on $tbW$ couplings \cite{cms17a}, as well as the very
recent limits on top dipole moments \cite{cms22} which were not
included in \cite{bou22}. We consider also the ATLAS collaboration
projections on limits on top dipole moments for the HL-LHC
\cite{ATLAS2018}.

A number of important issues discussed in \cite{bou22} are not covered
here for reasons of space. We point out among them, an extensive (and
hopefully exhaustive) analysis of background processes, a summary of
all global SMEFT top-quark analyses to date, and an in-depth
discussion of the computation of indirect limits on anomalous
couplings based on $B\to X_s\gamma$. We refer the reader to
\cite{bou22} for a comprehensive treatment of those topics.

This note is organized as follows. In section \ref{sec:eff.op} we
discuss the dimension-six SMEFT basis operators relevant to this
work. In section \ref{sec:photo} we discuss the top-pair
photoproduction process in $pe^-$ collisions in the SM, and its Monte
Carlo simulation and computation. In section \ref{sec:res.coupl} we
present our limits on top anomalous effective couplings, and compare
them to those obtained and projected by experimental
collaborations. Finally, in section \ref{sec:finrem}, we give our
final remarks.

\section{Effective SM Lagrangian}
\label{sec:eff.op}

The effective Lagrangian for the SM extended by dimension-six
gauge-invariant operators is of the form,
\begin{equation}
  \label{eq:lag}
  \mathcal{L} = \mathcal{L}_\mathrm{SM} + \frac{1}{\Lambda^2}\sum_{\mathcal{O}}
  (\widehat{C}_\mathcal{O} \mathcal{O}+\mathrm{h.c.})+\cdots, 
\end{equation}
where $\mathcal{O}$ denotes the dimension-six effective operators,
$\Lambda$ is the new-physics scale, and the ellipsis refers to
higher-dimensional operators. It is understood in (\ref{eq:lag}) that
the addition of the Hermitian conjugate, denoted $+\mathrm{h.c.}$ in
the equation, is applicable only to non-Hermitian operators.
Throughout this paper we use the dimension-six effective operators
from the Warsaw operator basis \cite{grz10}. In particular, we
use the same sign convention for covariant derivatives as in
\cite{bou13b,grz10}, namely, $D_\mu=\partial_\mu + i e A_\mu$ for the
electromagnetic coupling.  However, we adopt the operator
normalization defined in \cite{zha14} (see also \cite{manohar1309}),
where a factor $y_t$ is attached to an operator for each Higgs field
it contains, and a factor $g$ ($g'$) for each $W_{\mu\nu}$
($B_{\mu\nu}$) field-strength tensor.  The Wilson coefficients in
(\ref{eq:lag}) are denoted $\widehat{C}$, since we will denote $C$ the
coefficients associated with the original operator basis \cite{grz10}.
In fact, it will be convenient in what follows to express our results
in terms of the modified dimensionless couplings
\begin{equation}
  \label{eq:coupl}
  \C_\mathcal{O} = \widehat{C}_\mathcal{O} \frac{v^2}{\Lambda^2},
\end{equation}
where $v$ is the Higgs-field vacuum expectation value.  At tree level
the coupling constants $\C_\mathcal{O}$ are independent of the scale
$\Lambda$. We denote complex couplings as
$\C_{\mathcal{O}}=\C_{\mathcal{O}\,r}+i \C_{\mathcal{O}\,i}$.

There are seven operators in the basis \cite{grz10} that couple
electroweak bosons and third family quarks. One of them,
$Q^{33}_{\varphi u}$, does not contribute to the photoproduction
process. Other three of them, $Q^{33}_{\varphi ud}$, $Q^{33}_{uW}$,
$Q^{33}_{dW}$, are strongly limited by $W$-helicity fractions in
$t\bar{t}$ production and decay at the LHC and HL-LHC, while the
expected sensitivity of the top-pair photoproduction process to them
is expected to be low. For these reasons, we do not consider them
further. Of the three operators remaining, one linear combination,
$Q^{(+)33}_{\varphi q}=Q^{(3)33}_{\varphi q}+Q^{(1)33}_{\varphi q}$,
is strongly constrained by $b$-physics at LEP and SLAC
electron--positron colliders.  Therefore, we focus our analysis on the
two operators $\O_{uB}^{33}$ and $\O^{(-)33}_{\vphi q}$. Expanding
these operators in physical fields yields, with the conventions
discussed above:
\end{multicols}
\begin{wequation}
  \begin{aligned}
\O^{33}_{uB} &= y_t g' Q^{33}_{uB}= \sqrt{2} y_t e (v+h) (\partial_\mu
A_\nu - \tan\theta_W \partial_\mu Z_\nu)\; \bar{t}_{L}\sigma^{\mu\nu}
t_{R}~,\\    
\O_{\vphi q}^{(-)33} &= \O_{\vphi q}^{(3)33}-\O_{\vphi q}^{(1)33} =
-y_t^2 Q_{\vphi q}^{(-)33} \\
&=-y_t^2\frac{g}{\sqrt{2}} (v+h)^2 \left( W^+_\mu\, \bar{t}_{L}
  \gamma^\mu b_L +
  W^-_\mu\,\bar{b}_{L} \gamma^\mu t_L \right) - y_t^2\frac{g}{c_W} (v+h)^2 Z_\mu\, \bar{t}_{L}\gamma^\mu t_{L} \, ,\\
  \end{aligned}
  \label{eq:operators}
\end{wequation}
\begin{multicols}{2}
\noindent where $Q^{33}_{uB}$, $Q_{\vphi q}^{(-)33}$ are the basis
operators defined in \cite{grz10}.  Notice that both operators
$\O^{33}_{uB}$ and $\O_{\vphi q}^{(-)33}$ are $O(g^1)$ with respect to
the weak coupling constant, which makes the definitions
(\ref{eq:operators}) consistent from the point of view of perturbation
theory.  We stress here the definition $\O_{\vphi q}^{(-)33} =
\O_{\vphi q}^{(3)33}-\O_{\vphi q}^{(1)33}$ we use, since sometimes in
the literature the opposite sign is used. The effective Lagrangian
used throughout this paper results from substituting
(\ref{eq:operators}) and (\ref{eq:coupl}) in the Lagrangian
(\ref{eq:lag}).  It is convenient to record here the relation between
the Wilson coefficients in the form (\ref{eq:coupl}) and those
associated with the original basis \cite{grz10} (see also
\cite{newcite}), 
\begin{equation}
  \label{eq:coupl2}
  \begin{aligned}
  C_{uB}^{33} &= \frac{\Lambda^2}{v^2} y_t g'\cub = 5.906 \cub~,\\
  C_{\varphi q}^{(-)33} &= -\frac{\Lambda^2}{v^2} y_t^2 \cmpq = -16.495
  \cmpq~.     
  \end{aligned}
\end{equation}
The numerical values in this equation arise from the parameters
$\Lambda=1$ TeV, $v=246$ GeV, $g'=0.358$, $g=0.648$, $y_t=1$.

It is common practice in the literature to write the anomalous
interactions in terms of form factors.  We adopt here the definition
of top electromagnetic dipole moments given in eq.\ (2) of
\cite{bou13b}, and the CC vertex form factors from eq.\ (7.1) of
\cite{cms17a}. Comparing those equations (which are summarized in eq.\
(5) of \cite{bou22}), with (\ref{eq:lag}), (\ref{eq:coupl}),
(\ref{eq:operators}), yields the tree-level relations,
\begin{equation}
  \label{eq:relations}
    \kappa = 2 y_t^2 \widetilde{C}_{uB\,r}^{33},
  \qquad
  \widetilde{\kappa} = 2 y_t^2 \widetilde{C}_{uB\,i}^{33},
  \qquad
    \delta f_V^L = y_t^2 \widetilde{C}_{\vphi q}^{(-)33}~.
\end{equation}
These particularly simple relations are a consequence of eq.\
(\ref{eq:coupl}) and the operator normalization discussed in the text
immediately above that equation. We see from (\ref{eq:relations}), in
particular, that for all practical purposes $\delta f_V^L
=\widetilde{C}_{\vphi q}^{(-)33}$. 
\end{multicols}

\begin{figure*}[t]
  \centering{}
  \includegraphics[scale=0.7]{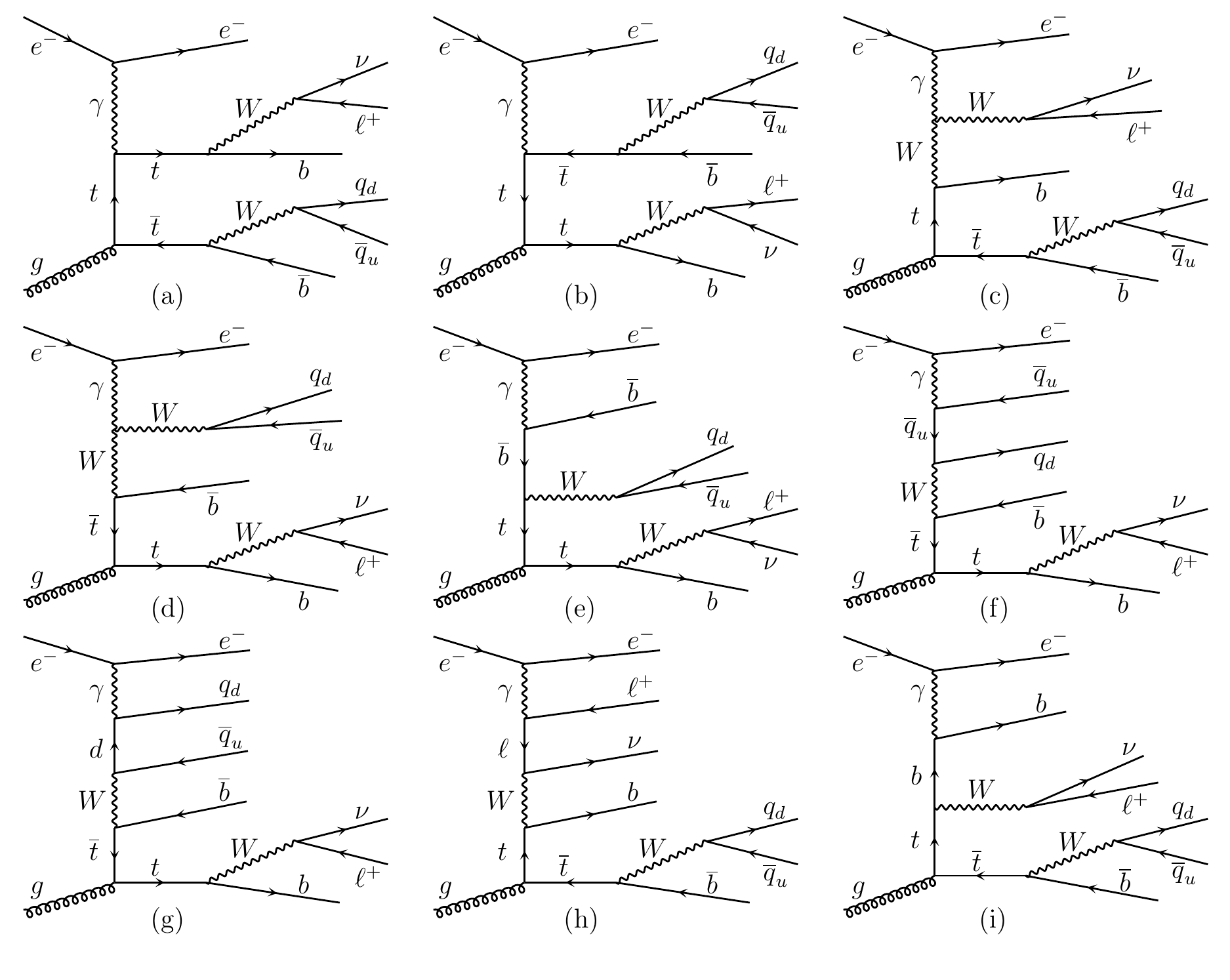}
  \caption{Unitary-gauge Feynman diagrams for the photoproduction of a
    top pair in semileptonic mode, see eq.\ (\ref{eq:ttx.proc}).
    \mbox{All~diagrams} for the final state
    $e^-\, b\ell^+\nu_\ell\, \bar{b}\bar{q}_uq_d$ are shown.  Diagrams
    (c)--(i) are necessary to preserve electromagnetic gauge
    invariance when $t$, $W$ are off shell.}
  \label{fig:feyn.1}
\end{figure*}

\begin{multicols}{2}
\section{Top-pair photoproduction in the SM}
\label{sec:photo}
We are interested in top-pair photoproduction in $pe^-$ collisions in
the semileptonic decay channel which, at parton level, leads to the
seven-fermion final states,
\begin{equation}
  \label{eq:ttx.proc}
    g\; e^- \rightarrow e^- t\bar{t}\rightarrow
    e^-\,b\ell^+ \nu_\ell\, \overline{b}\overline{q}_u q_d +
    e^-\, bq_u \overline{q}_d\,\overline{b}\ell^- \overline{\nu}_\ell,
\end{equation}
with $q_u=u,c,$ $q_d=d,s,$ $\ell=e,\mu$.  The set of Feynman diagrams
for this process in the photoproduction region in unitary gauge in the
SM with Cabibbo mixing is shown in figure \ref{fig:feyn.1}.  We
consider the top-pair photoproduction process defined by the diagrams
in that figure our signal process.

We can divide the set of diagrams in figure \ref{fig:feyn.1} into two
subsets: the first subset includes diagrams (a), (b), containing three
internal top lines, and the second one comprises the remaining
diagrams, (c)--(i), containing two internal top lines. The second
subset is necessary to preserve electromagnetic gauge invariance in
the phase-space regions where $t$ or $W$ lines are off shell. There
is, in fact, strong destructive interference between the two subsets
in the photoproduction region, such that the total cross section
computed with all the diagrams is smaller than the cross sections
obtained from (a), (b) or (c)--(i) separately, by a factor 10--25
depending on the cuts used in the computation.

We must consider also other processes with the same final state as
(\ref{eq:ttx.proc}), which constitute irreducible backgrounds.
Particularly important is the associate $tbW$ photoproduction, in
which $bW$ does not originate in a top decay. The Feynman diagrams for
$tbW$ production are given in figure 2 of \cite{bou22}. This process turns
out to be the dominant background to the signal process
(\ref{eq:ttx.proc}). Several other irreducible and reducible
backgrounds are discussed in sections 3 and 5 of \cite{bou22}.

We compute the tree-level cross section for top-pair photoproduction
and its backgrounds with MadGraph5\_aMC@NLO version 2.6.3
\cite{alw14}, together with Pythia version 6.428 \cite{sjo06} and
Delphes version 3.4.2 \cite{fav14}. The parameters of the simulation
are described in section 4 of \cite{bou22}, as are the details of the
top-reconstruction method, and the event-selection cuts. We are led to
define three photoproduction regions, characterized by the rapidity of
the scattered electron,
\begin{equation}
  \label{eq:php}
  \begin{aligned}
  PhP_{I}:   &\quad  -4.741< y(e^-) < -3.0,\\
  PhP_{II}:  &\quad  -5.435< y(e^-) < -3.0,\\
  PhP_{III}: &\quad  -6.215< y(e^-) < -3.0.
\end{aligned}
\end{equation}
As discussed in detail in \cite{bou22}, the sensitivity to
$\widetilde{C}_{uB}^{33}$ is highest in $PhP_{I}$ and lowest in
$PhP_{III}$, and the sensitivity to $\widetilde{C}_{\vphi q}^{(-)33}$
is highest in $PhP_{III}$ and lowest in $PhP_{I}$. Both sensitivities
are intermediate in $PhP_{II}$.

With the phase-space cuts specified in section 4 of \cite{bou22}, the
cross sections for the $t\bar{t}$ photoproduction signal process
(\ref{eq:ttx.proc}), figure \ref{fig:feyn.1}, and the $tbW$
irreducible background, are found to be as follows,
\setlength{\arraycolsep}{1.75pt}
\begin{equation}
  \label{eq:sm.xsctn.numeric}
  \begin{array}{c|ccc|ccc|}
             &\multicolumn{3}{c|}{\text{LHeC}}&\multicolumn{3}{c|}{\text{FCC-he}}\\
{[}\text{fb}]  &PhP_{I}&PhP_{II}&PhP_{III}&PhP_{I}&PhP_{II}&PhP_{III}\\\hline
t\bar{t} &0.40 &0.73 &1.32&4.28&6.19&10.51\\
tbW      &0.041&0.083&0.16&0.44&0.71&1.42
  \end{array}~,
\end{equation}
expressed in femtobarns.  We notice here that the $tbW$ background has
cross section at the parton level that is roughly 20\% of the signal
cross section at the LHeC, and roughly 35\% at the FCC-he, the precise
number depending on the photoproduction region. We designed the
phase-space cuts to reduce this background to levels below 15\%.  As
seen in (\ref{eq:sm.xsctn.numeric}), the $tbW$ background is 10\% of
the signal in region $PhP_{I}$ and 11.3\% in $PhP_{II}$ at both the
LHeC and FCC-he. In region $PhP_{III}$ we have 12.2\% at the LHeC and
13.5\% at the FCC-he.  This $tbW$ background proves to be the most
difficult one to control.

\section{Results for effective couplings}
\label{sec:res.coupl}

In this section we summarize the main results for the
effective-coupling limits from \cite{bou22}.

\subsection{Bounds on $\boldsymbol{\cmpq}$}
\label{sec:cmpq.results}

The largest sensitivity to $\cmpq$ is obtained in region $PhP_{III}$.
Indeed, the anomalous coupling $\cmpq$ constitutes a perturbation
$\delta f_V^L$ to the SM charged-current coupling
$f_V^L=1+\delta f_V^L$ and, therefore, it also perturbs the
cancellation among diagrams discussed above in section
\ref{sec:photo}.  Thus, the sensitivity is largest in region
$PhP_{III}$ where the cancellation is strongest.  We obtain limits for
$\cmpq$ at the LHeC and FCC-he energies, in photoproduction region
III, at one- and two-sigma levels, assuming a measurement uncertainty
of 12\%, \setlength{\arraycolsep}{5pt}
\begin{equation}
  \label{eq:cmpq.1}
\begin{array}{cc}
68\%\,\mathrm{C.L.}:&-0.039 < \delta f_V^L < 0.035\,,\\[3pt]
95\%\,\mathrm{C.L.}:&-0.083 < \delta f_V^L < 0.067~.
\end{array}
\end{equation}
These limits are obtained from a single total cross section value,
with no other observable involved. We express them in terms of $\delta
f_V^L$ to compare them to the limits reported by CMS; from fig.\ 6 of 
\cite{cms17a} we get,
\begin{equation}
  \label{eq:cmpq.2}  
\begin{array}{cc}
68\%\,\mathrm{C.L.}:&-0.024 < \delta f_V^L < 0.094\,,\\[3pt]
95\%\,\mathrm{C.L.}:&-0.062 < \delta f_V^L < 0.132~.
\end{array}
\end{equation}
By taking interval length as a measure of sensitivity, we see that
both limits in (\ref{eq:cmpq.1}) are significantly stronger than those
in (\ref{eq:cmpq.2}).

\subsection{Bounds on $\boldsymbol{\cub}$: single-coupling bounds}
\label{sec:cub.results.1}

The largest sensitivity to $\cub$ is obtained in region $PhP_{I}$.
This is due to the fact that the SM is close to an infrared divergence
at $Q^2=0$ and, therefore, as $Q^2$ decreases the SM cross section
grows much faster than the dipolar cross section, which is infrared
finite.  This causes the sensitivity to both $\cubr$, $\cubi$ to
decrease as we go from $PhP_{I}$ to $PhP_{III}$. We obtain limits on
$\cub$ at the LHeC and FCC-he energies, in photoproduction region I,
at one- and two-sigma levels, assuming a measurement uncertainty of
12\%,
\begin{equation}
  \label{eq:cub.lim}
\begin{array}{cc}
  68\%\,\mathrm{C.L.}: & \left\{
                         \begin{array}{c}
                       -0.24 < \ccubr < 0.29\,, \\
                       -0.89 < \ccubi < 0.89\,,
                         \end{array}
  \right.\\[5pt]
  95\%\,\mathrm{C.L.}: & \left\{
                         \begin{array}{c}
                           -0.45 < \ccubr < 0.65\,, \\
                           -1.24 < \ccubi < 1.24\,.
                         \end{array}
  \right.
\end{array}
\end{equation}
The limits on $\cubr$, which is proportional to the magnetic dipole
moment, are asymmetric because of the interference with the SM. Also
for that reason, they are stronger than those on the imaginary part.
Since the electric dipole moment operator is $CP$ odd, the
interference with the SM is very small, and the limits on $\cubi$ are
symmetric.

We compare the limits (\ref{eq:cub.lim}) with those projected for the
HL-LHC by the ATLAS collaboration \cite{ATLAS2018}, from the radiative
top-pair production and decay process $pp\to t\bar{t}\gamma$,
\begin{equation}
  \label{eq:hl.lhc.cub}
    95\%\,\mathrm{C.L.}: \qquad -0.5 < \ccubr < 0.3~.
\end{equation}
These are somewhat stricter than those we obtain, (\ref{eq:cub.lim}),
based on interval length.  We remark, however, that the ATLAS
projections are based on two channels ($\ell\ell$ and $\ell j$), and
involve the total cross section and two differential cross sections
each one spanning about six bins. There are, in total, about a dozen
measurements involved in the limits (\ref{eq:hl.lhc.cub}), whereas
(\ref{eq:cub.lim}) are based on a single observable, the total cross
section.

The limits set on $\ccub$ by the ATLAS and CMS collaborations from
measurements of $t\bar{t}\gamma$ production are nowadays incorporated
into global analyses; we refer to \cite{bou22} for a detailed review
of those. Very recently, the CMS collaboration \cite{cms22} has
measured the total cross section for $pp\to t\bar{t}\gamma$, as well
as two differential cross sections ($d\sigma/dp_{T\gamma}$,
$d\sigma/d\eta_\gamma$), in two reaction channels ($\ell\ell$ and
$\ell j$). The limits obtained from the dilepton channel,
\begin{equation}
  \label{eq:cms.1}
  95\%\,\mathrm{C.L.}: \qquad \left\{
    \begin{array}{c}
      -1.08 < \ccubr < 1.10\,, \\
      -1.08 < \ccubi < 1.21\,, 
    \end{array}
    \right.
\end{equation}
are significantly weaker than ours, (\ref{eq:cub.lim}), by interval
length. The limits obtained from a combination of both channels are
reported to be,
\begin{equation}
  \label{eq:cms.2}
  95\%\,\mathrm{C.L.}: \qquad \left\{
    \begin{array}{c}
      -0.64 < \ccubr < 0.75\,, \\
      -0.75 < \ccubi < 0.79\,,
    \end{array}
    \right.
\end{equation}
and are substantially stronger than for each separate channel. The
limits (\ref{eq:cms.2}) are only slightly weaker than
(\ref{eq:cub.lim}) for $\cubr$, but definitely stronger for $\cubi$.
As is the case for the limits (\ref{eq:hl.lhc.cub}), the strong limits
(\ref{eq:cms.2}) are the result of combining more than a dozen
observables: the total cross section and two differential cross
sections, for two reaction channels.

\subsection{Bounds on $\boldsymbol{\cub}$: allowed two-coupling regions}
\label{sec:cub.results.2}

\begin{figure*}[t]
  \centering{}
  \includegraphics[scale=0.8]{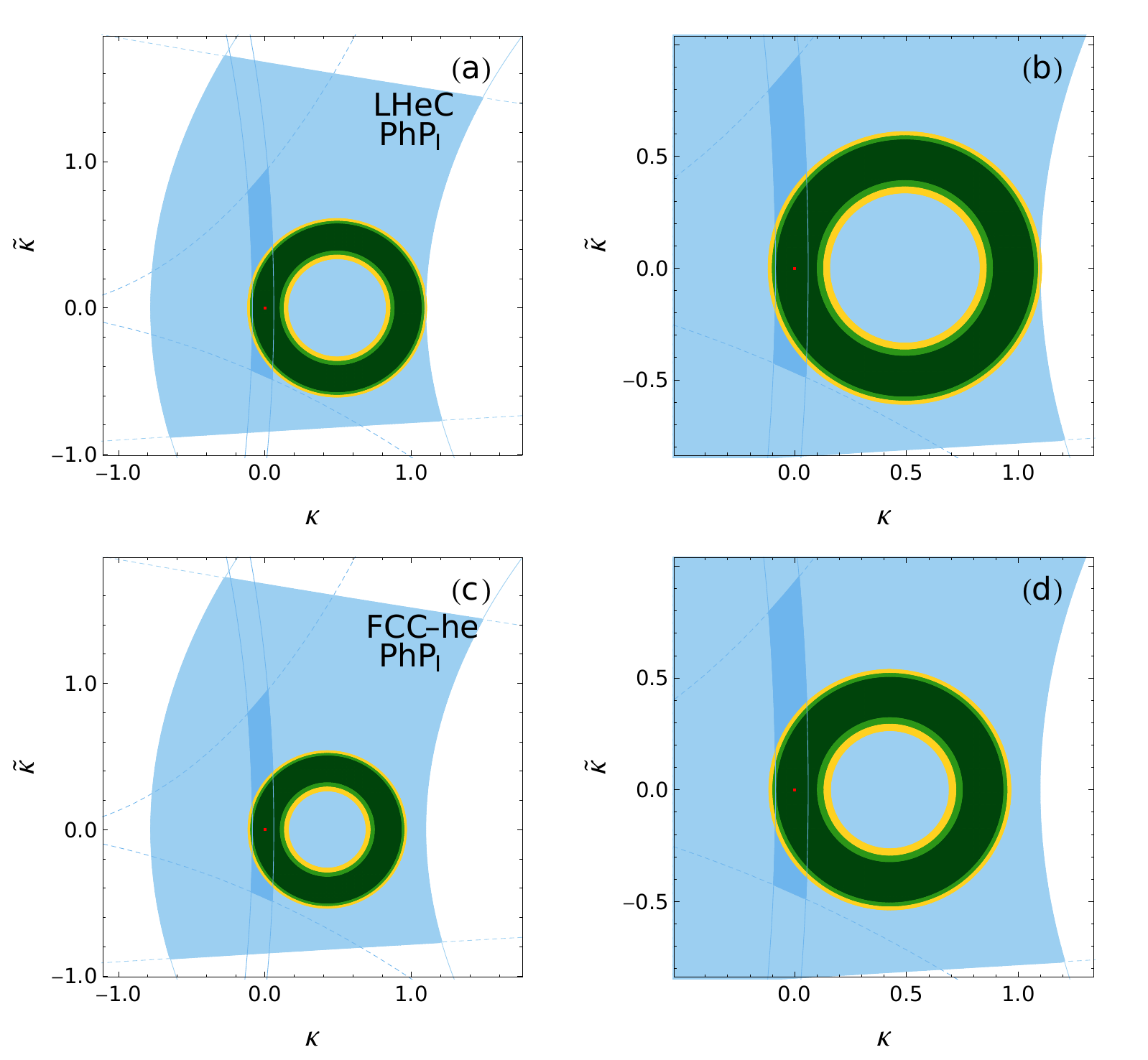}
  \caption{Allowed regions for the
      top quark dipole moments $\kappa$ and $\widetilde{\kappa}$ at
      (a),(b) the LHeC and (c),(d) the FCC-he. Panels (a), (c) display
      a global view, (b), (d) a magnified one.  Annular regions:
      regions allowed at 68\% C.L.\ by a top-pair
      tagged-photoproduction cross-section measurement, in
      photoproduction region $PhP_{I}$ (\ref{eq:php}), with
      experimental uncertainties 12\% (dark green), 15\% (light
      green), and 18\% (yellow). Light-blue
      area: region allowed by the measurements of the branching ratio
      and $CP$ asymmetry of $B\rightarrow X_s\gamma$ decays, using the
      results from \cite{hewett}.  Darker-blue
      area: same as previous, but using the results from
      \cite{crivellin15}.} 
  \label{fig:1}
\end{figure*}

In figure \ref{fig:1} we show the allowed regions in the
$\kappa$--$\widetilde{\kappa}$ plane, determined by the top-pair
photoproduction cross section at both the LHeC and FCC-he energies, in
region $PhP_{I}$ at 68\% C.L. These can be related to $\cub$ through
(\ref{eq:relations}), and to $C_{uB}^{33}$ through (\ref{eq:coupl2}).
The allowed regions, given by the circular coronas, correspond to the
assumed measurement uncertainties $\veps_\mathrm{exp}=12$, 15, 18\% in
different colors as indicated in the figure caption.
Also seen in figure \ref{fig:1} is that the annular allowed regions
obtained at the FCC-he are somewhat smaller than those at the LHeC
energy.  We notice, however, that both sets of allowed regions are
identical in the neighborhood of the origin (i.e., the SM), which is
consistent with the individual-coupling bounds we obtain being the
same at both energies.  

Also shown in the figure are the regions in the
$\kappa$--$\widetilde{\kappa}$ plane allowed by the branching ratio
and $CP$ asymmetry for the process $B\rightarrow X_s\gamma$, in both
the form obtained from \cite{hewett}, and the form from
\cite{crivellin15}. The difference in area between these two regions
hardly needs to be emphasized. We remark, however, that even the
smaller region resulting from \cite{crivellin15} is not completely
contained in the annular regions determined by top-pair
photoproduction, which results in a significant reduction of the
allowed parameter space.
\end{multicols}

\begin{multicols}{2}

\section{Final remarks}
\label{sec:finrem}

In this note we summarize the results of our analysis
\cite{bou22} of top-pair photoproduction in semileptonic mode in $pe$
collisions at the LHeC and FCC-he. Our main results are the limits
(\ref{eq:cmpq.1}) on $\cmpq$ ($\delta f_V^L$), and those on $\cubr$,
$\cubi$ ($\kappa$, $\widetilde{\kappa}$), (\ref{eq:cub.lim}), and the
two-dimensional allowed regions for $\kappa$, $\widetilde{\kappa}$ in
figure \ref{fig:1}. We also made a detailed comparison of our results
to those from ATLAS and CMS from \cite{cms17a,ATLAS2018,cms22}.

Based on our results, we expect the LHeC to provide limits on $\cmpq$
($\delta f_V^L$) similar to those from the HL-LHC. We also expect the
LHeC to obtain limits on $\cub$ stronger than those from the
HL-LHC. These will then be the strongest until the operation of
$e^-e^+$ colliders begins.  Both sets of limits will constitute an
important contribution to future global analyses. The FCC-he can yield
improved sensitivity to $\cmpq$ and $\cub$, relative to that of the
LHeC, from substantially larger statistics and improved systematics.

The recent strong results on limits on top electromagnetic dipole
moments by the LHC collaborations \cite{ATLAS2018,cms22} suggest,
however, that we should upgrade our analysis of top photoproduction in
$pe$ colliders by including appropriate differential cross
sections. An enhanced background rejection from a more robust
multivariate analysis would also help strengthen the bounds on top
couplings. These extensions to our analysis are currently in progress;
we will report the results elsewhere.

\end{multicols}
\end{document}